\documentclass[showpacs,nofootinbib]{revtex4}
\usepackage{amsmath}
\usepackage{amssymb}
\usepackage{epsfig}

\newcommand{\La}{{\Lambda}}
\newcommand{\Si}{{\Sigma}}
\newcommand{\be}{\begin{eqnarray}}
\newcommand{\ee}{\end{eqnarray}}

\begin{document} 
\title{To bind or not to bind: The $H$-dibaryon in light of chiral effective field theory}

\author{Johann Haidenbauer$^{1}$ and Ulf-G. Mei{\ss}ner$^{1,2}$}

\affiliation{
$^1$Institute for Advanced Simulation, Institut f\"ur Kernphysik (Theorie) and J\"ulich Center for 
Hadron Physics, Forschungszentrum J\"ulich, D-52425 J\"ulich, Germany \\
$^2$Helmholtz-Institut f\"ur Strahlen- und Kernphysik (Theorie)
and Bethe Center for Theoretical Physics,
Universit\"at Bonn, Nu\ss allee 14-16, D-53115 Bonn, Germany
}

\begin{abstract}
We analyse the quark mass dependence of the binding energy of the $H$-dibaryon
in the framework of chiral effective field theory. We show that the SU(3)
breaking effects induced by the differences of the pertinent two-baryon
thresholds ($\La\La$, $\Xi N$, $\Si\Si$) 
have a very pronounced impact that need to be incorporated properly
in future lattice QCD simulations. We also point out that if the
$H$-dibaryon is a two-baryon bound state, its dominant component is $\Xi N$
rather than $\La\La$, which is a consequence of the approximate SU(3)
flavor symmetry of the two-baryon interactions.  
\end{abstract}
\pacs{13.75.Ev,12.39.Fe;14.20.Pt}

\maketitle

\section{Introduction and summary}

In 1977 Jaffe predicted a deeply bound 6-quark state with $I=0$, $J=0$, ${S}=-2$
from the bag-model, called the $H$-dibaryon \cite{Jaffe:1976yi}. 
Subsequently, many experimental searches for the $H$-dibaryon were carried out,
but so far no convincing signal was found \cite{Yoon}.
However, recently evidence for a bound $H$-dibaryon was claimed based on
lattice QCD calculations \cite{Beane,Inoue,Inoue11a,Beane11a}. Extrapolations of the
simulations, performed for pion masses $M_\pi \gtrsim 400$ MeV, to the physical mass
suggest that the $H$-dibaryon could be either loosely bound or move into
the continuum \cite{Beane11,Shanahan11}.

In this paper, we analyze various issues related to these lattice results
in the framework of chiral effective field theory (EFT) for the baryon-baryon ($BB$) 
interaction at leading order (LO) in the Weinberg counting. This scheme
has proven successful for the few data on hyperon-nucleon scattering \cite{Polinder06}
and also for the bounds that exist on the $BB$ interactions with
strangeness $S=-2$ \cite{Polinder07}. There is one low-energy constant (LEC)
in the $S=-2$ sector, corresponding to the SU(3) flavor-singlet channel, 
that can not be fitted by hyperon-nucleon data and can be
fine-tuned to produce a bound $H$ with a given binding energy. This framework
also allows us to study the quark mass dependence\footnote{Because of the
Gell-Mann-Oakes-Renner relation, the pion mass squared is proportional
to the average light quark mass. Therefore, the notions ``quark mass dependence''
and ``pion mass dependence'' can be used synonymously.} of the binding energy of the
$H$, see the calculations of the quark mass dependence of the deuteron
binding energy in \cite{Beane:2002vs,Beane03,Epe02,Epe02a}. 
Another important issue to be addressed here is how this quark mass 
dependence is affected when the SU(3) breaking manifested in the masses
of the octet baryons is accounted for.
In the real world the $BB$ thresholds are not degenerate but 
they all differ for the relevant $\Lambda\Lambda$, $\Sigma\Sigma$ 
and $\Xi N$ systems and that has very definite dynamical consequences, 
as we will show in what follows.

The pertinent results of our study can be summarized as follows:
\begin{itemize}
\item[(i)] We have analyzed the effective range expansion for the 
$\Lambda\Lambda$ $^1S_0$ channel, assuming a loosely bound $H$ dibaryon.
It shows a very different behaviour to the case of the deuteron in
the neutron-proton $^3S_1$ channel. In fact, any attraction supplied 
by the flavor-singlet channel contributes with a much larger weight 
to the $\Xi N$ interaction than to $\Lambda\Lambda$, according to SU(3)
flavor symmetry, so that the $H$-dibaryon should predominantly be a 
$\Xi N$ bound state.
\item[(ii)] 
We observe that, for pion masses below 400~MeV, the dependence of the binding
energy of the $H$ is linearly decreasing with decreasing pion mass, in
agreement with the findings in Ref.~\cite{Beane11}. In particular, a $H$
binding energy (BE) adjusted to the value found by NPLQCD \cite{Beane} 
at $M_\pi = 389\,$MeV, is reduced by 7~MeV at the physical pion mass. 
For larger pion masses, this dependence is weakened. Note, however, that
for such large pion masses this should only be considered a trend as 
the chiral EFT is constructed for masses/momenta well below the chiral
symmetry breaking scale.
\item[(iii)] We find a much more drastic effect caused by
the SU(3) breaking related to the
values of the three thresholds $\Lambda\Lambda$, $\Sigma\Sigma$ and $\Xi N$.
For physical values the BE of the $H$ is reduced by as much as 60~MeV as 
compared to a calculation based on degenerate (i.e. SU(3) symmetric) $BB$ 
thresholds. 
Translating this observation to the situation in the HAL QCD \cite{Inoue} 
calculation, we see that the bound state has disappeared at the physical point. 
For the case of the NPLQCD 
calculation, a resonance in the $\Lambda\Lambda$ system might survive.
\end{itemize}

Our manuscript is organized as follows: In Sec.~\ref{sec:2}, we recall the
basic formalism of the $BB$ interaction in the framework of chiral
EFT. Sec.~\ref{sec:3} contains a detailed discussion of the quark mass
dependence of the BE of the $H$ and the influence of the SU(3) breaking
through the various two-baryon thresholds. In Sec.~\ref{sec:4} we try to 
make direct contact to the results published by the NPLQCD and HAL QCD
collaborations.

\section{The baryon-baryon interaction to leading order}
\label{sec:2}

For details on the derivation of the chiral $BB$ potentials for the strangeness sector 
at LO using the Weinberg power counting, we refer the reader to Refs.~\cite{Polinder06,Polinder07},
see also Refs.~\cite{Savage1,Korpa,Savage2}.
Here, we just briefly summarize the pertinent ingredients of the chiral EFT for $BB$ interactions. 
\begin{table*}[t]
\renewcommand{\arraystretch}{1.2}
\centering
\begin{tabular}{|l|c|c|l|c|l|}
\hline
&Channel &Isospin &$C_{1S0}$ &Isospin &$C_{3S1}$\\
\hline
$S=0$&$NN\rightarrow NN$ &$1$ & $C^{27}$ &$0$ &$C^{10^*}$\\
\hline
$S=-1$&$\Lambda N \rightarrow \Lambda N$ &$\frac{1}{2}$ &$\frac{1}{10}\left(9C^{27}+C^{8_s}\right)$
&$\frac{1}{2}$ &$\frac{1}{2}\left(C^{8_a}+C^{10^*}\right)$\\
&$\Lambda N \rightarrow \Sigma N$ &$\frac{1}{2}$ &$\frac{3}{10}\left(-C^{27}+C^{8_s}\right)$
&$\frac{1}{2}$ &$\frac{1}{2}\left(-C^{8_a}+C^{10^*}\right)$\\
&$\Sigma N \rightarrow \Sigma N$ &$\frac{1}{2}$ &$\frac{1}{10}\left(C^{27}+9C^{8_s}\right)$
&$\frac{1}{2}$ &$\frac{1}{2}\left(C^{8_a}+C^{10^*}\right)$\\
&$\Sigma N \rightarrow \Sigma N$ &$\frac{3}{2}$ &$C^{27}$
&$\frac{3}{2}$ &$C^{10}$\\
\hline
$S=-2$&$\Lambda\Lambda \rightarrow \Lambda\Lambda$ &$0$ & $\frac{1}{40}\left(27C^{27}+8C^{8_s}+5C^{1}\right)$
  & & \\
&$\Lambda\Lambda \rightarrow \Xi N$ &$0$ &$\frac{-1}{40}\left(18C^{27}-8C^{8_s}-10\,C^{1}\right)$
  & & \\
&$\Lambda\Lambda \rightarrow \Sigma\Sigma$ &$0$ &$\frac{\sqrt{3}}{40}\left(-3C^{27}+8C^{8_s}-5C^{1}\right)$
  & & \\
&$\Xi N \rightarrow \Xi N$ &$0$ &$\frac{1}{40}\left(12C^{27}+8C^{8_s}+20\,C^{1}\right)$
  &$0$ &$C^{8_a}$\\
&$\Xi N \rightarrow \Sigma\Sigma$ &$0$ &$\frac{\sqrt{3}}{40}\left(2C^{27}+8C^{8_s}-10\,C^{1}\right)$
  &$1$ &$\frac{\sqrt{2}}{6}\left(C^{10}+C^{10^*}-2C^{8_a}\right)$\\
&$\Sigma\Sigma \rightarrow \Sigma\Sigma$ &$0$ &$\frac{1}{40}\left(C^{27}+24C^{8_s}+15C^{1}\right)$
 &$1$ &$\frac{1}{6}\left(C^{10}+C^{10^*}+4C^{8_a}\right)$\\
&$\Xi N \rightarrow \Xi N$ &$1$ &$\frac{1}{5}\left(2C^{27}+3C^{8_s}\right)$
  &$1$ &$\frac{1}{3}\left(C^{10}+C^{10^*}+C^{8_a}\right)$\\
&$\Xi N \rightarrow \Sigma\Lambda$ &$1$ &$\frac{\sqrt{6}}{5}\left(C^{27}-C^{8_s}\right)$
  &$1$ &$\frac{\sqrt{6}}{6}\left(C^{10}-C^{10^*}\right)$\\
&$\Sigma\Lambda \rightarrow \Sigma\Lambda$ &$1$ &$\frac{1}{5}\left(3C^{27}+2C^{8_s}\right)$
  &$1$ &$\frac{1}{2}\left(C^{10}+C^{10^*}\right)$\\
&$\Sigma\Lambda \rightarrow \Sigma\Sigma$ &   &   &$1$ &$\frac{\sqrt{3}}{6}\left(C^{10}-C^{10^*}\right)$\\
&$\Sigma\Sigma \rightarrow \Sigma\Sigma$ &$2$ &$C^{27}$
  & & \\
\hline
\end{tabular}
\caption{Various LO baryon-baryon contact potentials for the ${}^1S_0$ and ${}^3S_1$ partial
waves in the isospin basis. $C^{27}$ etc. refers to the corresponding ${\rm SU(3)_f}$
irreducible representation.}
\label{tab:1}
\end{table*}
\renewcommand{\arraystretch}{1.0}

The LO potential consists of four-baryon contact terms without derivatives and of 
one-pseudoscalar-meson exchanges. 
 The LO ${\rm SU(3)}_{\rm f}$ invariant contact terms for the octet $BB$
interactions that are Hermitian 
and invariant under Lorentz transformations follow from the Lagrangians
\begin{eqnarray}
{\mathcal L}^1 &=& C^1_i \left<\bar{B}_a\bar{B}_b\left(\Gamma_i B\right)_b\left(\Gamma_i B\right)_a\right>\ , \quad
{\mathcal L}^2 = C^2_i \left<\bar{B}_a\left(\Gamma_i B\right)_a\bar{B}_b\left(\Gamma_i B\right)_b\right>\ , \nonumber \\
{\mathcal L}^3 &=& C^3_i \left<\bar{B}_a\left(\Gamma_i B\right)_a\right>\left<\bar{B}_b\left(\Gamma_i B\right)_b\right>\  .
\label{eq:2.1}
\end{eqnarray}
Here $a$ and $b$ denote the Dirac indices of the particles, $B$ is the irreducible octet (matrix) 
representation of ${\rm SU(3)}_{\rm f}$, and the $\Gamma_i$ are the usual elements of the 
Clifford algebra \cite{Polinder06}. As described in Ref.~\cite{Polinder06}, 
to LO the Lagrangians in Eq.~(\ref{eq:2.1}) give rise to only six independent 
low-energy coefficients (LECs), the $C_i^j$s in Eq.~(\ref{eq:2.1}), due to 
${\rm SU(3)}_{\rm f}$ constraints. They need to be determined by a fit to experimental data. 
It is convenient to re-express the $BB$ potentials in terms of the ${\rm SU(3)_f}$ 
irreducible representations, see e.g. Refs.~\cite{Swart,Dover}.
Then the contact interaction is given by
\begin{equation}
V=
\frac{1}{4}(1-\mbox{\boldmath $\sigma$}_1\cdot \mbox{\boldmath $\sigma$}_2) \, C_{1S0}
+ \frac{1}{4}(3+\mbox{\boldmath $\sigma$}_1 \cdot\mbox{\boldmath $\sigma$}_2) \, C_{3S1} \ ,
\label{contact}
\end{equation}
and the constraints imposed by the assumed ${\rm SU(3)}_{\rm f}$ symmetry on the interactions
in the various $BB$ channels for the $^1S_0$ and $^3S_1$ partial waves can be
readily read off from Table~\ref{tab:1}.
 
The lowest order ${\rm SU(3)}_{\rm f}$ invariant pseudoscalar-meson--baryon
interaction Lagrangian embodying the appropriate symmetries was also discussed in \cite{Polinder06}. 
The invariance under ${\rm SU(3)}_{\rm f}$ 
transformations implies specific relations between the various coupling constants, namely
\begin{equation}
\begin{array}{rlrlrl}
f_{NN\pi}  = & f, & f_{NN\eta_8}  = & \frac{1}{\sqrt{3}}(4\alpha -1)f, & f_{\Lambda NK} = & -\frac{1}{\sqrt{3}}(1+2\alpha)f, \\
f_{\Xi\Xi\pi}  = & -(1-2\alpha)f, &  f_{\Xi\Xi\eta_8}  = & -\frac{1}{\sqrt{3}}(1+2\alpha )f, & f_{\Xi\Lambda K} = & \frac{1}{\sqrt{3}}(4\alpha-1)f, \\
f_{\Lambda\Sigma\pi}  = & \frac{2}{\sqrt{3}}(1-\alpha)f, & f_{\Sigma\Sigma\eta_8}  = & \frac{2}{\sqrt{3}}(1-\alpha )f, & f_{\Sigma NK} = & (1-2\alpha)f, \\
f_{\Sigma\Sigma\pi}  = & 2\alpha f, &  f_{\Lambda\Lambda\eta_8}  = & -\frac{2}{\sqrt{3}}(1-\alpha )f, & f_{\Xi\Sigma K} = & -f.
\end{array}
\label{su3}
\end{equation}
Here $f\equiv g_A/2F_\pi$, where $g_A$ is the nucleon axial-vector strength
and $F_\pi$ is the weak pion 
decay constant.  We use the values $g_A= 1.26$ and $F_\pi = 92.4$~MeV.
For $\alpha$, the $F/(F+D)$-ratio \cite{Polinder06}, we adopt 
the SU(6) value: $\alpha=0.4$, which is consistent with recent determinations
of the axial-vector coupling constants \cite{Ratcliffe}.

The spin-space part of the LO one-pseudoscalar-meson-exchange potential is similar to the 
static one-pion-exchange potential in chiral EFT for nucleon-nucleon
interactions, see e.g. \cite{Epe98} (recoil and relativistic corrections give 
higher order contributions),
\begin{eqnarray}
V^{B_1B_2\to B_1'B_2'}&=&-f_{B_1B_1'P}f_{B_2B_2'P}\frac{\left(\mbox{\boldmath $\sigma$}_1\cdot{\bf q}\right)\left(\mbox{\boldmath $\sigma$}_2\cdot{\bf q}\right)}{{\bf q}^2+M^2_P}\ ,
\label{eq:14}
\end{eqnarray}
where $M_P$ is the mass of the exchanged pseudoscalar meson. The transferred 
and average momentum, ${\bf q}$ and ${\bf k}$, are defined in terms of the final and initial 
center-of-mass (c.m.) momenta of the baryons, ${\bf p}'$ and ${\bf p}$, as 
${\bf q}={\bf p}'-{\bf p}$ and ${\bf k}=({\bf p}'+{\bf p})/2$. 
In the calculation we use the physical masses of the exchanged pseudoscalar mesons. 
The explicit ${\rm SU(3)}$ breaking reflected in the mass splitting between the 
pseudoscalar mesons and, in particular, the small mass of the pion relative to
the other members of the octet leads to sizeable differences in the range of
the interactions in the different channels and, thus, induces an essential dynamical 
breaking of ${\rm SU(3)}$ symmetry in the $BB$ interactions.
The $\eta$ meson was identified with the octet $\eta$ ($\eta_8$) and its physical 
mass was used.

The reaction amplitudes are obtained from the solution of a coupled-channels 
Lippmann-Schwinger (LS) equation for the interaction potentials: 
\begin{eqnarray}
&&T_{\rho''\rho'}^{\nu''\nu',J}(p'',p';\sqrt{s})=V_{\rho''\rho'}^{\nu''\nu',J}(p'',p')+
\sum_{\rho,\nu}\int_0^\infty \frac{dpp^2}{(2\pi)^3} \, V_{\rho''\rho}^{\nu''\nu,J}(p'',p)
\frac{2\mu_{\nu}}{q_{\nu}^2-p^2+i\eta}T_{\rho\rho'}^{\nu\nu',J}(p,p';\sqrt{s})\ .
\label{LS} 
\end{eqnarray}
The label $\nu$ indicates the particle channels and the label $\rho$ the partial wave. 
$\mu_\nu$ is the pertinent reduced mass. The on-shell momentum in the intermediate state, 
$q_{\nu}$, is defined by $\sqrt{s}=\sqrt{m^2_{B_{1,\nu}}+q_{\nu}^2}+\sqrt{m^2_{B_{2,\nu}}+q_{\nu}^2}$. 
Relativistic kinematics is used for relating the laboratory energy $T_{{\rm lab}}$ of the hyperons 
to the c.m. momentum.

In \cite{Polinder06,Polinder07} 
the LS equation was solved in the particle basis, in order to incorporate the correct physical
thresholds. Since here we are only interested in the $H$ dibaryon we work in the isospin
basis. Then for $J=0$, $I=0$, and $S=-2$ we have to consider the three coupled channels
$\La\La$, $\Xi N$ and $\Si\Si$. We use the following (isospin averaged) masses 
$m_\La=1115.6$ MeV, $m_\Si=1192.5$ MeV, $m_\Xi=1318.1$ MeV, and $m_N=939.6$ MeV so
that the $\La\La$, $\Xi N$, and $\Si\Si$ thresholds are at
2231.2,  2257.7, and 2385.0~MeV, respectively. Furthermore, the potentials in the LS 
equation are cut off with a regulator function, $\exp\left[-\left(p'^4+p^4\right)/\Lambda^4\right]$, 
in order to remove high-energy components of the baryon and pseudoscalar meson fields \cite{Epe05}.
We consider cut-off values in the range 550, ..., 700 MeV, similar to what was used for  
chiral $NN$ potentials \cite{Epe05}.

The imposed ${\rm SU(3)}$ flavor symmetry implies that only five of the six LECs 
contribute to the $YN$ interaction, namely $C^{27}$, $C^{10}$, $C^{10^*}$, $C^{8_s}$, 
and $C^{8_a}$, cf. Table~\ref{tab:1}. 
These five contact terms were determined in 
\cite{Polinder06} by a fit to the $YN$ scattering data. Since the $NN$ data
cannot be described with a LO EFT, ${\rm SU(3)}$ constraints from the $NN$ interaction 
were not implemented explicitly. As shown in Ref.~\cite{Polinder06},  a good
description  of the 35 low-energy $YN$ scattering can be 
obtained for cutoff values $\Lambda=550,...,700$ MeV and for natural values of the LECs. 
The sixth LEC ($C^{1}$) is only present in the $S=-2$ channels with isospin zero,
cf. Table~\ref{tab:1}. There is scarce experimental information on these 
channels that could be used to fix this LEC, but it turned out that the quality 
of the existing data do not really allow to constrain its value reliably 
\cite{Polinder07}. Even with the value of the sixth LEC chosen so that 
$C^{\La\La\to \La\La}_{1S0} = 0$, agreement with those data can be achieved. In this case 
a scattering length of $a_{^1S_0}^{\Lambda\Lambda} = -1.52$~fm \cite{Polinder07} 
is obtained. 
Analyses of the measured binding energy of the double-strange hypernucleus
${}^{\;\;\;6}_{\Lambda\Lambda}{\rm He}$ \cite{Takahashi:2001nm} suggest that
the $\La\La$ scattering length could be in the range of 
$-1.3$ to $-0.7$ fm \cite{Gal,Rijken,Fujiwara}.
 
\section{Quark mass dependence of the binding energy and SU(3) breaking effects}
\label{sec:3}

Chiral effective field theory itself does not allow one to make any
predictions with regard to the existence of the $H$ dibaryon because, as stated,
one of the contact terms (${C^1}$) occurs just in the channel in question, cf. 
Table~\ref{tab:1}, and has to be determined from there. 
If we take over the values for $C^{27}$ and $C^{8_s}$ as fixed from the $Y N$ data 
and assume that ${C^1} = 0$ then we find no bound state for $\Lambda\Lambda$ - 
neither in LO \cite{Polinder06} nor based on the preliminary NLO results \cite{Hai10}. 
However, if we assume that ${C^1} \neq 0$ and vary its value then 
in both cases a near-threshold bound state can be produced for values of natural size.

At the same time, the framework of chiral effective field theory in which our 
$\Lambda\Lambda$ interaction is derived is very well suited to shed light 
on the general characteristics of a $H$-dibaryon, should it indeed exist.
In particular, it allows us study the implications of the imposed 
(approximate) SU(3) flavor symmetry and to explore the dependence of 
the properties of an assumed $H$-dibaryon on
the masses of the relevant mesons and baryons. The latter aspect is a 
rather crucial issue in view of the fact that the available lattice QCD 
calculations were not performed at the physical masses of the involved 
particles. 

To start our discussion, let us assume that the $H$-dibaryon is a 
(loosely) bound $BB$ state 
and that its binding energy $E_H$ is similar to that of
the deuteron $D$. For pedagogical purposes we fix the value of the flavor-singlet LEC 
${C^1}$ in such a way that
$\gamma_{H} = \gamma_{D} = 0.23161\,$fm
($E = - \gamma^2 / m_{B}$, where $m_{B}$ is either $m_N$ or $m_\Lambda$),
because of the well-known relation between the binding energy and the effective
range parameters \cite{Schwinger47,Bethe} 
\begin{equation}
\frac{1}{a} \simeq {\gamma} - \frac{1}{2}{r} {\gamma}^2 .
\nonumber
\end{equation}
This relation is very well fulfilled for the deuteron and the corresponding
neutron-proton $^3S_1$ scattering length ($a=5.43\,$fm) and effective range
($r=1.76\,$fm). One would naively expect that the same should happen for the
$H$-dibaryon. However, it turns out that the corresponding results for
$\La\La$ in the $^1S_0$ partial wave are quite different, namely
$a=3.00\,$fm and $r=-4.98\,$fm. Specifically, the effective range
is much larger and, moreover, negative. Clearly, the
properties of the $H$-dibaryon are not comparable to those of the
deuteron, despite the fact that both bound states are close to the
elastic threshold.
Indeed, if one recalls the expressions for the relevant potentials
as given in Table \ref{tab:1},
\begin{equation}
V^{\La\La \rightarrow \La\La} = \frac{1}{40}\left(27C^{27}+8C^{8_s}+{ 5}{
    C^{1}}\right) \, , \ \ \ 
V^{\Xi N \rightarrow \Xi N} = \frac{1}{40}\left(12C^{27}+8C^{8_s}+{ 20}{
    C^{1}}\right) \, ,
\nonumber 
\end{equation}
one can see that the attraction supplied by the SU(3) flavor-singlet state ($C^{1}$)
contributes with a much larger weight to the $\Xi N$ channel than to $\La\La$.
This indicates that the presumed $H$-dibaryon could be predominantly a
$\Xi N$ bound state. We have confirmed this conjecture by evaluating explicitly
the phase shifts in the $\La\La$ and $\Xi N$ channels, cf. the discussion in the
next section. Indeed, one finds that the phase shift for the $\Xi N$ channel
is rather similar to the $NN$ $^3S_1$ case. Specifically, the $\Xi N$ ($^1S_0$) phase shift 
$\delta(q_{\Xi N})$ fulfills $\delta(0)-\delta(\infty) = 180^\circ$ in 
agreement with the Levinson theorem. 
The $\La\La$ ($^1S_0$) phase behaves rather differently and 
satisfies $\delta(q_{\La\La}=0) - \delta(\infty) = 0$.
Note that there have been earlier discussions on this issue in
the context of $S=-2$ baryon-baryon interactions derived within the
quark model \cite{Oka,Nakamoto}. 

\begin{figure}[t!]
\centering
\includegraphics[height=9.cm,keepaspectratio,angle=-90]{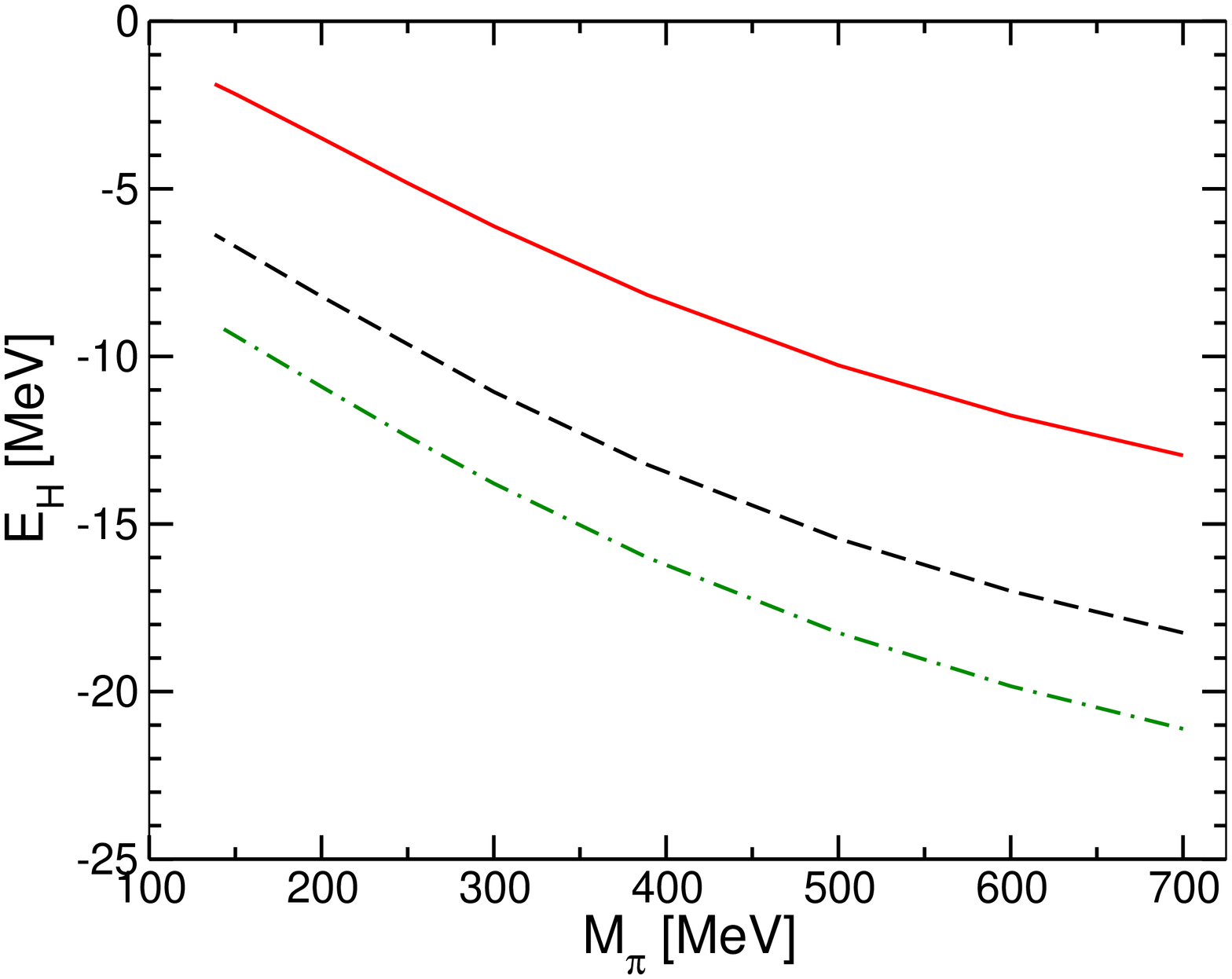}\includegraphics[height=9.cm,keepaspectratio,angle=-90]{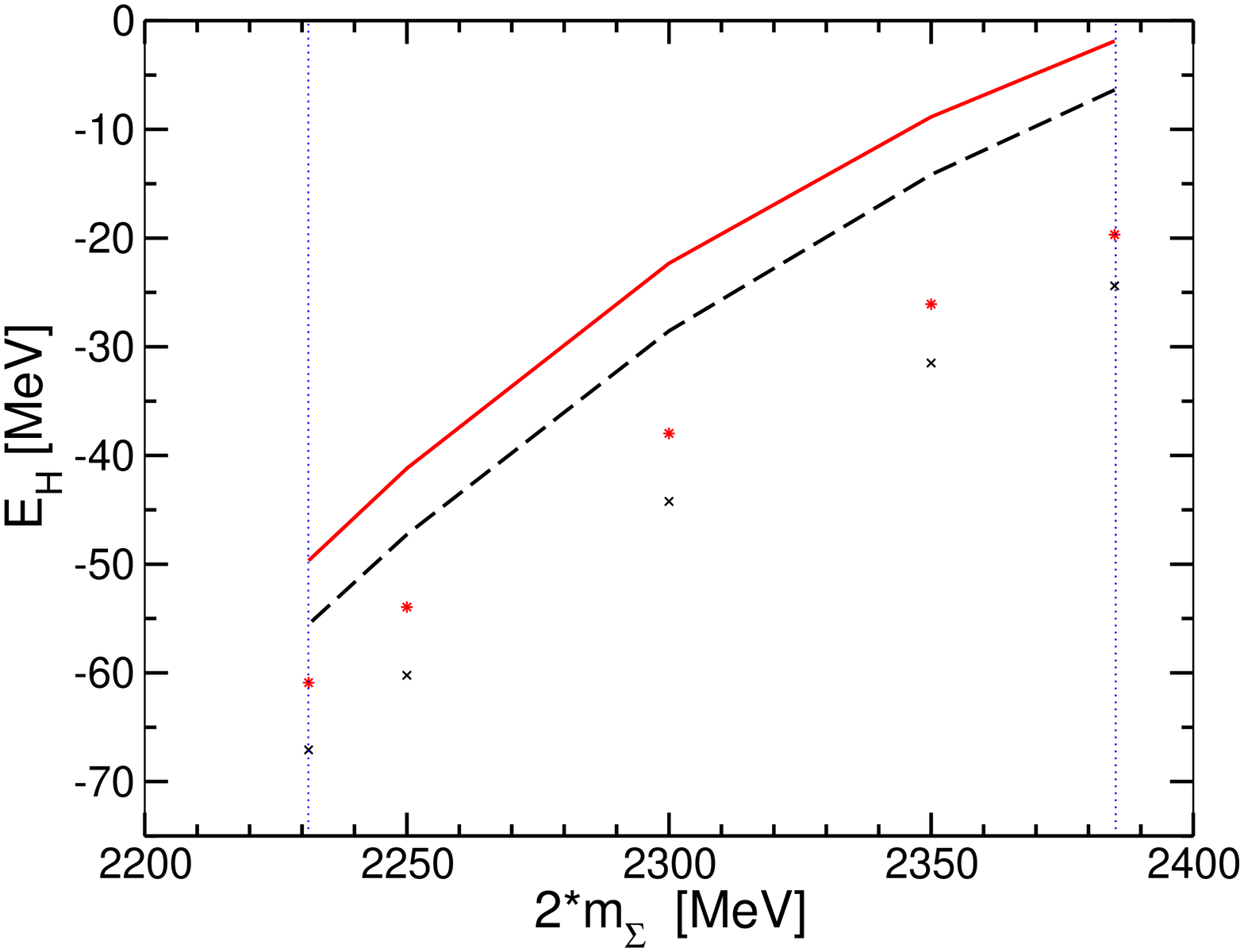}
\caption{Dependence of the binding energy of the $H$-dibaryon on the pion mass 
$M_\pi$ (left) and on the $\Sigma$ mass $m_\Sigma$ (right).
The solid curve correspond to the case where $C^1$ is fixed such that $E_H = -1.87\,$MeV 
for physical masses while for the dashed curve $C^1$ is fixed to yield $E_H = -13.2\,$MeV for 
$M_\pi = 389$ MeV. The pion mass dependence is also shown for $E_H = -16.6\,$MeV 
(dash-dotted curve).
The asterisks and crosses represent results where, besides the 
variation of $m_\Sigma$, $m_\Xi+m_N = 2 m_\La$ is assumed so that the $\Xi N$ threshold 
coincides with that of the $\La\La$ channel. 
The vertical (dotted) lines indicate the physical $\La\La$ and $\Si\Si$ thresholds. 
}
\label{fig:mpi}
\end{figure}

Let us now consider variations of the masses of the involved particles. 
The dependence of the $H$ binding energy on the pion mass $M_\pi$ is 
displayed in Fig.~\ref{fig:mpi} (left). For the case considered above, 
enlarging the pion mass to around 400 MeV 
(i.e. a value corresponding to the NPLQCD calculation \cite{Beane}) increases the
binding energy to around 8 MeV and a further change of $M_\pi$ to 700 MeV
(corresponding roughly to the HAL QCD calculation \cite{Inoue}) yields then 13 MeV, 
cf. the solid line. 
Readjusting $C^1$ so that we predict a $H$ binding energy of 13.2 MeV for
$M_\pi=389$ MeV, corresponding to the latest result published by 
NPLQCD~\cite{Beane11a}, yields the dashed curve. 
Fig.~\ref{fig:mpi} includes also results of a calculation where $C^1$ was 
fixed in order to reproduce their earlier value of around 16 MeV \cite{Beane} 
(dash-dotted curve), to facilitate a more direct comparison with
chiral extrapolations~\cite{Beane11,Shanahan11} based on that value.
It is obvious that the dependence on $M_\pi$ we get agrees well -- at least on a 
qualitative level -- with that presented in Ref.~\cite{Beane11}. Specifically, our 
calculation exhibits the same trend (a decrease of the binding energy with 
decreasing pion mass) and our binding energy of 9 MeV at the physical 
pion mass is within the error bars of the results given in~\cite{Beane11}. 
On the other hand, we clearly observe a non-linear dependence of the binding
energy on the pion mass. As a consequence, scaling our results to the binding
energy reported by the HAL QCD collaboration \cite{Inoue} (30-40 MeV for  
$M_\pi \approx 700-1000$~MeV) yields binding energies of more than 20 MeV
at the physical point, which is certainly outside of the range suggested in
Ref.~\cite{Beane11}. However, we note that for such large pion masses the LO
chiral EFT can not be trusted quantitatively. 
We remark that in our simulations the curves corresponding to different binding 
energies remained roughly parallel even up to such large values as suggested 
by the HAL QCD collaboration. 

Our finding that any $H$-dibaryon is very likely a bound $\Xi N$ state rather than a $\La\La$
state, which follows from the assumed (approximate) SU(3) symmetry of the interaction, 
suggests that not only the pion mass
but also the masses of the baryons play a significant role for the concrete
value of binding energy. 
Indeed, the physical difference between the $\La\La$ and $\Xi N$ thresholds of around
26~MeV implies that the $H$-dibaryon considered above is, in reality, bound by roughly
28~MeV with respect to its ``proper'' threshold. Accordingly, one intuitively expects that 
in a fully SU(3) symmetric case, where the masses of all octet baryons coincide, the 
bound state would remain basically fixed to the $\Xi N$ threshold and then would lie also 
about 28 MeV below the $\La\La$ threshold. In the concrete case of $J=0$, $I=0$, $S=-2$ we are 
dealing with three coupled channels, namely $\La\La$, $\Xi N$, and $\Si\Si$. 
Since we know from our experience with coupled-channel problems 
\cite{Polinder06,Hai05,Hai10a,Hai11} that coupling effects are sizeable and 
the actual separation of the various thresholds plays a crucial role we investigated also
the dependence of the $H$ binding energy on the thresholds (i.e. on the $\Sigma$, and on 
the $\Xi$ and $N$ masses). Corresponding results are displayed in Fig.~\ref{fig:mpi} on
the right side. The solid curve is again the result based on our reference case 
with a binding energy of -1.87~MeV for physical masses of the pion and the baryons. 
When we now decrease the $\Sigma$ mass so that its nominal threshold of 2385~MeV
moves downwards and finally coincides with the one of the $\La\La$ channel (2231.2~MeV),
we observe a rather drastic change in the $H$ binding energy. Note that the direct
interaction in the $\Sigma\Sigma$ channel is actually repulsive for the low-energy
coefficients $C^{27}$ and $C^{8_s}$ fixed from the $YN$ data plus the pseudoscalar
meson exchange contributions with coupling constants determined from the SU(3)
relations Eq.~(\ref{su3}). And it remains repulsive even for $C^{1}$ values that produce
a bound $H$-dibaryon. But the coupling between the channels generates a sizeable
effective attraction which increases when the channel thresholds come closer. 
The dashed curve is a calculation with the contact term $C^{1}$ fixed to simulate 
the binding energy (-13.2~MeV) of the NPLQCD collaboration at $M_\pi=389\,$MeV.
As one can see, the dependence on the binding energy on the $\Sigma$ mass is rather
similar. The curve is simply shifted downwards by around 4.5~MeV, i.e. by the difference
in the binding energy observed already at the physical masses. 
The asterisks and crosses represent results where, besides the variation of the 
$\Sigma\Sigma$ threshold, the $\Xi N$ threshold is shifted to coincide with that
of the $\La\La$ channel. This produces an additional increase of the $H$ binding 
energy by 20~MeV at the physical $\Sigma\Sigma$ threshold and by 9~MeV 
for that case where all three $BB$ threshold coincide. 
Altogether there is an increase in the binding energy of roughly 60~MeV 
when going from the physical point to the case of baryons with identical
masses. This is significantly larger than the variations due to the pion mass
considered before. Note that we have kept the pion mass at its physical value
while varying the $BB$ thresholds. 

\section{Application to lattice QCD results}
\label{sec:4}
 
After these exemplary studies let us now try to connect with the published 
$H$ binding energies from the lattice QCD calculations \cite{Beane,Inoue}. 
The results obtained by the HAL QCD collaboration are obviously for the SU(3)
symmetric case and the corresponding masses are given in Table I of 
Ref.~\cite{Inoue}. Thus, we can take those masses and then fix the LEC
$C^1$ so that we reproduce their $H$ binding energy with those masses. To be
concrete, we use $m_{ps} = 673\,$MeV and $m_{B} = 1485\,$MeV, and fix 
$C^1$ so that $E_H = -35\,$MeV. When we now let the masses of the baryons and
mesons go to their physical values the bound state moves upwards, crosses
the $\La\La$ threshold, crosses also the $\Xi N$ threshold and then 
disappears. In fact, qualitatively this outcome
can be already read off from the curves in Fig.~\ref{fig:mpi} by combining the
effects from the variations in the pion and the baryon masses. Based on those
results one would expect a shift of the $H$ binding energy in the order of
60 to 70~MeV for the mass parameters of the HAL QCD calculation. 

In case of the NPLQCD calculation we take the values provided in 
Ref.~\cite{Beane11b}, i.e. $m_N=1151.3\,$MeV, $m_\La=1241.9\,$MeV, $m_\Si=1280.3\,$MeV,
and $m_\Xi=1349.6\,$MeV. Those yield
then 17~MeV for the $\Xi N$-$\La\La$ threshold separation (to be compared 
with the physical value of roughly 26~MeV) and 
77~MeV for the $\Si\Si$-$\La\La$ separation (physical value around 154~MeV). 
With those baryon masses we fix again the LEC $C^1$ so that we 
reproduce the $H$ binding energy given by the NPLQCD collaboration, namely 
$E_H =-13.2$~MeV \cite{Beane11a}. We use also $M_\pi = 389\,$MeV,
but we take the physical masses for the other pseudoscalar mesons ($K$ and $\eta$).
Again we let the masses of the baryons and of the pion go to their 
physical values. Also here the bound state moves upwards and crosses
the $\La\La$ threshold. However, in the NPLQCD case the state survives
and remains below the $\Xi N$ threshold at the physical point. Specifically,
we observe a resonance at a kinetic energy of 21~MeV in the $\La\La$
system or, more precisely, a quasi-bound state in the $\Xi N$ system around 
5~MeV below its threshold. 

\begin{figure}[t!]
\centering
\includegraphics[height=9.cm,keepaspectratio,angle=-90]{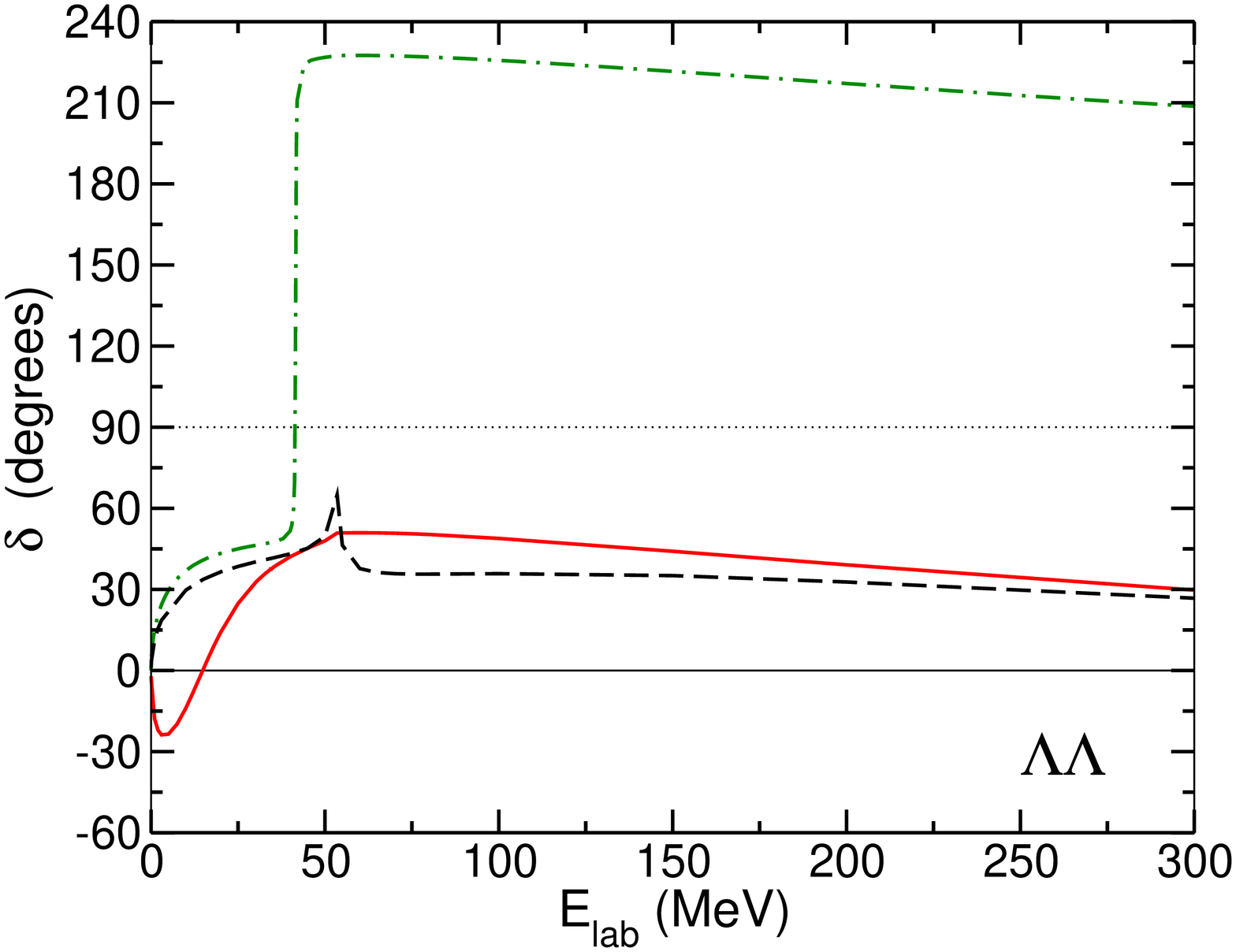}\includegraphics[height=9.cm,keepaspectratio,angle=-90]{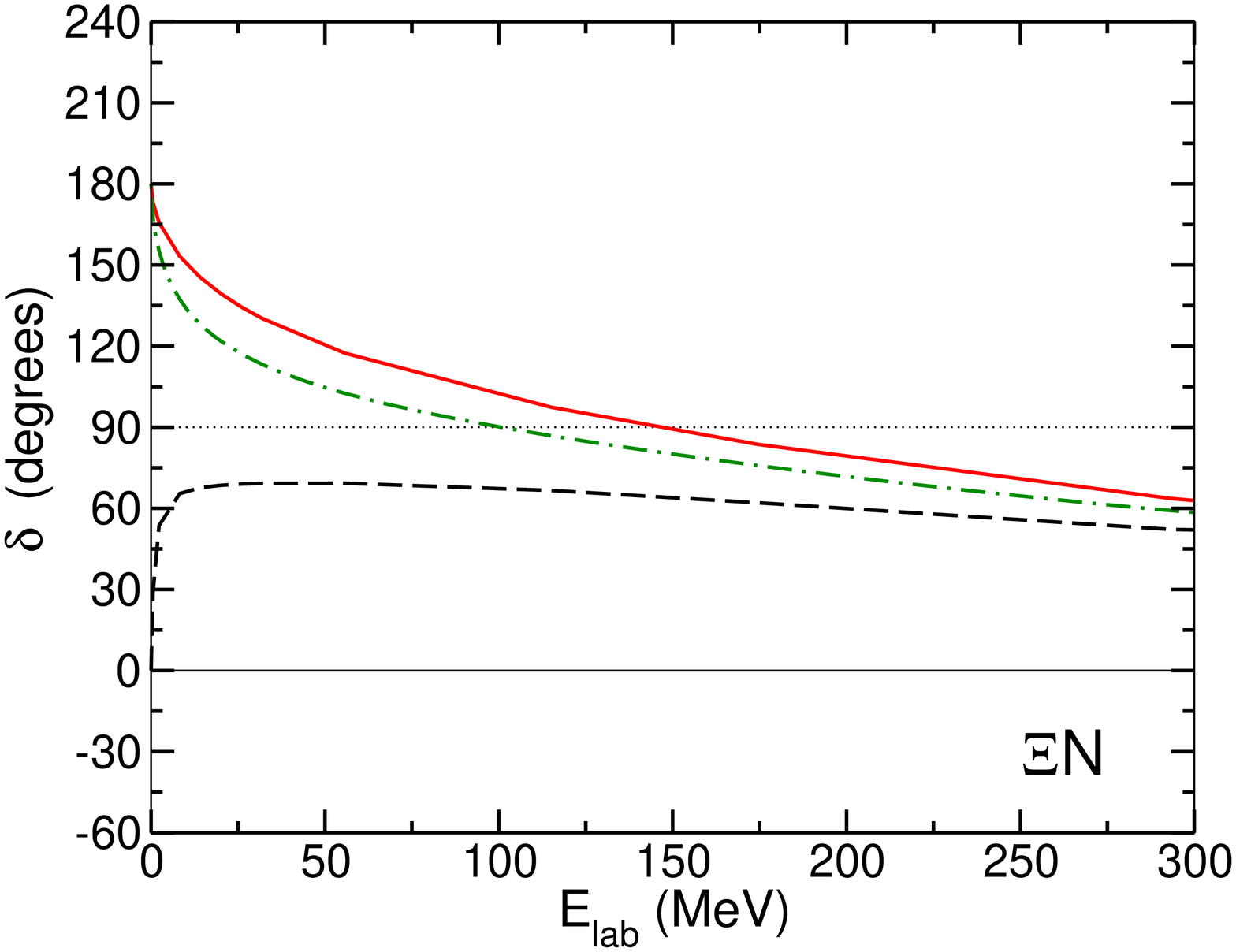}
\caption{Phase shifts for $\La\La$ ($^1S_0$) (left) and $\Xi N$ ($^1S_0$) (right)
as a function of the pertinent laboratory energies. 
The solid line is the result for our reference $BB$ interaction that produces a bound $H$ at
$E_H=-1.87$ MeV. The other curves are results for interactions that are fine-tuned to the 
$H$ binding energies found in the lattice QCD calculations of the HAL QCD (dashed) and 
NPLQCD (dash-dotted) collaborations, respectively, for the pertinent meson (pion) and
baryon masses as described in the text. 
}
\label{fig:phases}
\end{figure}

The results reported so far were all obtained with the LECs as fixed from
the $YN$ data for the cutoff value $\Lambda = 550$~MeV \cite{Polinder06}.
In order to investigate the stability of our results we 
considered also variations of the cutoff mass $\Lambda$ in the LS equation 
Eq.~(\ref{LS}) between 550 and 700~MeV, as in \cite{Polinder07},
and repeated the exercise described above. Those variations led to 
changes of the predicted resonance energy in the order of 1~MeV only. 
Even an exploratory calculation that utilizes the low-energy coefficients 
determined in a preliminary NLO study of the $YN$ system \cite{Hai10} 
yielded practically the same result. 

It is interesting to observe that the chiral extrapolation of the lattice
QCD results performed by Shanahan et al. \cite{Shanahan11} yields results
that are similar to ours. In that reference the authors conclude 
that the $H$-dibaryon is likely to be unbound by 13$\pm$14 MeV at the physical point. 
Let us emphasize, however, that our values are not really comparable
with theirs. In our analysis we assume that the $H$-dibaryon is actually 
a bound $BB$ state -- which seems to be the case also in the lattice QCD
studies \cite{Beane,Inoue}. On the other hand, in Ref.~\cite{Shanahan11} it is assumed that
the $H$ is a compact, multi-quark state rather than a loosely bound
molecular state. How such a genuine multi-quark state would be influenced
by variations of the $BB$ thresholds is completely unclear. It depends,
among other things, on whether and how strongly this state couples to the
$\La\La$, $\Xi N$, and $\Si\Si$ channels. So far there is no information
on this issue from lattice QCD calculations. Clearly, in case of a strong and 
predominant coupling to the $\La\La$ alone, variations of the $\Si\Si$ and
$\Xi N$ would not influence the $H$ binding energy significantly. However,
should it couple primarily to the $\Xi N$ and/or $\Si\Si$ channels
then we expect a sensitivity to their thresholds comparable to what we found in
our study for the case of a bound state. 

Finally, for illustrative purposes, let us show phase shifts for the
$^1S_0$ partial wave of the $\La\La$ and $\Xi N$ channels. This is done in
Fig.~\ref{fig:phases}. The solid line is the result for our reference $BB$ 
interaction that produces a loosely bound $H$ dibaryon with $E_H=-1.87$ MeV.
The phase shift for the $\Xi N$ channel (left side) is rather similar to the 
one for the $^3S_1$ $NN$ partial wave where the deuteron resides, 
see e.g. \cite{Epe05}. 
Specifically, it starts at $180^o$, decreases smoothly and eventually 
approaches zero for large energies, fulfilling the Levinson theorem. 
The result for $\La\La$ ($^1S_0$) (left side) behaves rather differently. 
This phase commences at zero degrees, is first negative but becomes positive within 20 MeV 
and finally turns to zero again for large energies.
The dashed curve corresponds to the interaction that was fitted to the result
of the HAL QCD collaboration and reproduces their bound $H$ dibaryon with
their meson and baryon masses. 
The phase shift of the $\Xi N$ channel, calculated with physical masses, 
shows no trace of a bound state anymore. Still the phase shift 
rises up to around $60^o$ near threshold, a behavior quite similar to that of 
the $^1S_0$ $NN$ partial wave where there is a virtual state (also called
antibound state \cite{Pearce}). Indeed, such a virtual state seems to 
be present too in the $\Xi N$ channel as a remnant of the original bound state.
The effect of this virtual state can be seen in the $\La\La$ phase shift where it
leads to an impressive cusp at the opening of the $\Xi N$ channel, 
cf. the dashed line on the left side. 
In the $\Xi N$ phase shifts for the NPLQCD case (dash-dotted curve) the
presence of a bound state is clearly visible. The corresponding $\La\La$
phase shift exhibits a resonance-like behavior at the energy where the
(quasi-bound) $H$ dibaryon is located.  

Given the present uncertainties of the lattice QCD results for the 
$H$ dibaryon binding energy we have focused here primarily on 
qualitative features of the $H$ that can be inferred from chiral 
effective field theory. Clearly, once more precise lattice data 
become available one should also perform a careful assessment of the
uncertainties involved in the EFT calculation.

\acknowledgments{
 We would like to thank S.~R.~Beane, A.~Gal, and M.~J.~Savage for their
constructive comments. 
 This work is supported by the Helmholtz Association by funds provided to
the virtual institute ``Spin and Strong QCD'' (VH-VI-231), by the EU-Research
Infrastructure Integrating Activity ``Study of Strongly Interacting Matter''
(HadronPhysics2, grant n. 227431) under the Seventh Framework Program of the EU,
and by the DFG (SFB/TR 16 ``Subnuclear Structure of Matter'').
}

\end{document}